\documentclass{article}

\usepackage{arxiv}

\usepackage[utf8]{inputenc} 
\usepackage[T1]{fontenc}    
\usepackage{hyperref}       
\usepackage{url}            
\usepackage{booktabs}       
\usepackage{amsfonts}       
\usepackage{nicefrac}       
\usepackage{microtype}      
\usepackage{lipsum}		
\usepackage{graphicx}
\usepackage[numbers]{natbib}
\usepackage{doi}

\usepackage{url}
\usepackage{tabulary}
\usepackage{svg}
\usepackage{soul}
\usepackage{amsmath,amssymb}
\usepackage[none]{hyphenat}
\usepackage{xurl} 
\usepackage{caption}
\usepackage{subcaption}
\usepackage{comment}
\usepackage{tablefootnote}
\usepackage{svg}
\usepackage{academicons}
\usepackage{bm}

\usepackage{bbding}

\graphicspath{{Figures/}}

\usepackage{url}
\usepackage{tabulary}
\usepackage{svg}
\usepackage{soul}
\usepackage{amsmath,amssymb}
\usepackage{amsfonts}
\usepackage{graphicx}
\usepackage[none]{hyphenat}
\usepackage[T1]{fontenc}
\usepackage{xurl} 
\usepackage{booktabs}
\usepackage{nicefrac}
\usepackage{caption}
\usepackage{subcaption}
\usepackage{hyperref}
\usepackage{comment}
\usepackage{tablefootnote}

\usepackage{svg}
\usepackage{academicons}
\usepackage{bm}

\usepackage{hyperref}

\usepackage{academicons}
\usepackage{xcolor,graphicx}
\usepackage{hyperref}

\definecolor{orcid-green}   {RGB} {166, 206, 57}

\usepackage[nolist,nohyperlinks]{acronym}

\title{Federated Learning: Organizational Opportunities, Challenges, and Adoption Strategies}

\author
{
Joaquin Delgado Fernandez\href{https://orcid.org/0000-0003-1326-6134}{\includegraphics[scale=0.06]{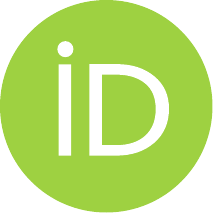}}$^{1,2}$, Martin Brennecke\href{https://orcid.org/0000-0002-1364-8119}{\includegraphics[scale=0.06]{orcid.pdf}}$^{1}$,
 Alexander Rieger\href{https://orcid.org/0000-0001-7996-4678}{\includegraphics[scale=0.06]{orcid.pdf}}$^{1,2}$, \\ \textbf{Tom Barbereau}\href{https://orcid.org/0000-0002-8554-0991}{\includegraphics[scale=0.06]{orcid.pdf}}$^{1}$, \textbf{Gilbert Fridgen}\href{https://orcid.org/0000-0001-7037-4807}{\includegraphics[scale=0.06]{orcid.pdf}}$^{1}$
\\
\normalsize{$^{1}$Interdisciplinary Centre for Security, Reliability and Trust, University of Luxembourg,}\\
\normalsize{29 Av. John F. Kennedy, Luxembourg, Luxembourg}
 \\
 \normalsize{$^{2}$Sam M. Walton College of Business, University of Arkansas,}\\
 \normalsize{1 University of Arkansas, Fayetteville, USA}
\\
 \normalsize{$^\ast$To whom correspondence should be addressed; E-mail: joaquin.delgadofernandez@uni.lu.}
}

\renewcommand{\shorttitle}{Federated Learning: Organizational Opportunities, Challenges, and Adoption Strategies}

\hypersetup{
pdftitle={Federated Learning: Organizational Opportunities, Challenges, and Adoption Strategies},
pdfsubject={cs.AI},
pdfauthor={Joaquin Delgado Fernandez, Martin Brennecke,Alexander Rieger, Tom Barbereau,  Gilbert Fridgen},
pdfkeywords={Artificial intelligence, Deep learning, Federated learning, Organization theory},
}

\begin{document}
\maketitle
\begin{abstract}
Restrictive rules for data sharing in many industries have led to the development of federated learning. Federated learning is a machine-learning technique that allows distributed clients to train models collaboratively without the need to share their respective training data with others. In this paper, we first explore the technical foundations of federated learning and its organizational opportunities. Second, we present a conceptual framework for the adoption of federated learning, mapping four types of organizations by their artificial intelligence capabilities and limits to data sharing. We then discuss why exemplary organizations in different contexts — including public authorities, financial service providers, manufacturing companies, as well as research and development consortia — might consider different approaches to federated learning. To conclude, we argue that federated learning presents organizational challenges with ample interdisciplinary opportunities for information systems researchers.
\end{abstract}

\keywords{Artificial intelligence \and Deep learning \and Federated learning \and Organization theory}

\sloppy

\acresetall

\begin{acronym}
\acro{ML}[ML]{Machine Learning}
\acro{DL}[DL]{Deep Learning}
\acro{FL}[FL]{Federated Learning}
\acro{AI}[AI]{Artificial Intelligence}
\acro{DP}[DP]{Differential Privacy}
\acro{SecAgg}[SecAgg]{secure aggregation}
\acro{SMPC}[SMPC]{secure multiparty computation}
\acro{IS}[IS]{Information Systems}
\acro{IT}[IT]{Information Technology}
\acro{FI}[FI]{financial institution}
\acro{DSR}[DSR]{Design Science Research}
\acro{Fed-Avg}[Fed-Avg]{Federated Average}
\acro{Fed-SGD}[Fed-SGD]{Federated Stochastic gradient descent}
\acro{GDPR}[GDPR]{General Data Protection Regulation}
\acro{NN}[NN]{neural networks}
\acro{EU}[EU]{European Union}
\acro{DLT}[DLT]{Distributed Ledger Technologies}
\acro{GPU}[GPU]{graphics processing unit}
\acro{IIoT}[IIoT]{Industrial Internet of Things}
\acro{SME}{small and medium-sized enterprise}
\acro{PdM}[PdM]{Predictive Maintenance}
\acro{DSR}{Design Science Research}
\acro{CSR}{Case Study Research}
\acro{GT}{Grounded Theory}
\acro{TOE}{Technology–Organization–Environment}
\acro{CCPA}{California Consumer Privacy Act}
\acro{DSO}{Distributor System Operator}
\acrodefplural{DSO}[DSOs]{Distributor Energy Operators}

\end{acronym}

\section{Introduction}
\label{sec:introduction}
Artificial intelligence (AI) capabilities\acused{AI} have become an important competitive advantage in various contexts \citep{berente2021managing}. Open-source models and \ac{AI}-as-a-service offerings make it cheap to acquire these capabilities, provided that organizations can train and operate the underlying models with the "right" data \citep{lins2021artificial, guntupalli2023aiasas}. However, accessing the "right" data both in terms of quantity and quality is not always possible due to competitive or privacy-related concerns \citep{agahari2022datasharing, quach2022privacydata}.

Federated learning (FL)\acused{FL} has the potential to facilitate better data access and overcome organizational data-sharing restrictions. It enables organizations to cooperate in training a shared \ac{ML} model without sharing data across organizational boundaries \citep{Kalra2023-sh}. By extension, it allows organizations in data-driven environments to co-create value from data without compromising data protection or losing their competitive edge \citep{bracht2023dataprotection, lacity1998empirical}. \ac{FL} thus has promising applications, ranging from financial services \citep{uberlee_2023} and healthcare \citep{kaissis2021end} to public administration \citep{Kalra2023-sh, sprenkampfederated, pati2022federated}, and the energy industry \citep{FERNANDEZ2022119915}. 

In this article, we explain the technical foundations of \ac{FL}, how it works at a conceptual level, and how it differs from conventional or centralized approaches to training \ac{ML} models. We then discuss constellations in which \ac{FL} can create organizational value before discussing the challenges that come with its adoption and implementation. Based on these adoption challenges, we present opportunities for further technical, organizational, and legal research on \ac{FL}.

\section{Technical Foundations}
\label{sec:tech}

Federated learning was born out of projects at Google and OpenAI. These projects sought to use data generated by keyboards on mobile devices to train \ac{ML} models that would enhance user experiences through next-word prediction. The project teams quickly realized that much of the collected data was highly personal and sensitive, which complicated its upload to company servers ~\citep{abadi2016deep,mcmahan2017communicationefficient}. \ac{FL} was then introduced in 2016 to circumvent these complications and enable a ``machine learning setting where the [...] training data remains distributed over a large number of clients'' \citep{konecny2016federated}. In the following years, the use of \ac{FL} was extended to other areas, not least due to the ability of FL to support \ac{ML} on data that is not independently and identically distributed (non-IID)~\citep{zhao2018federated}. More recent advances also eliminated the need for synchronous training and communication \citep{xu2021asynchronous} or central servers for coordination of the decentralized learning process \citep[see][]{Kalra2023-sh,circular_FL,shen2020federated}.

\begin{figure}[ht]
     \centering
     \begin{subfigure}[b]{0.45\textwidth}
         \centering
         \includegraphics[scale=0.6]{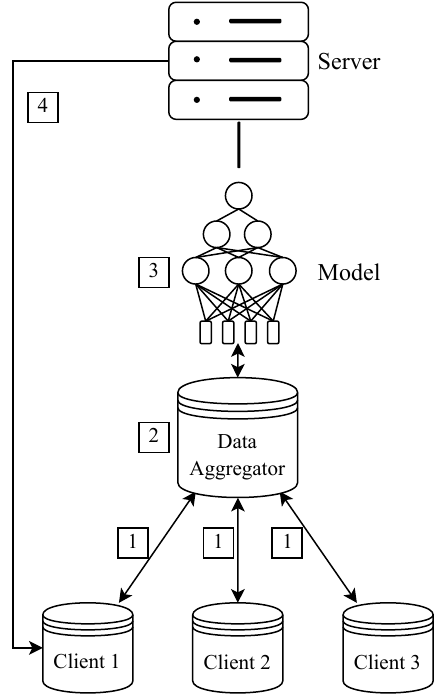}
         \caption{Centralized.}
         \label{fig:centralized}
     \end{subfigure}
     \hfill
     \begin{subfigure}[b]{0.45\textwidth}
         \centering
         \includegraphics[scale=0.6]{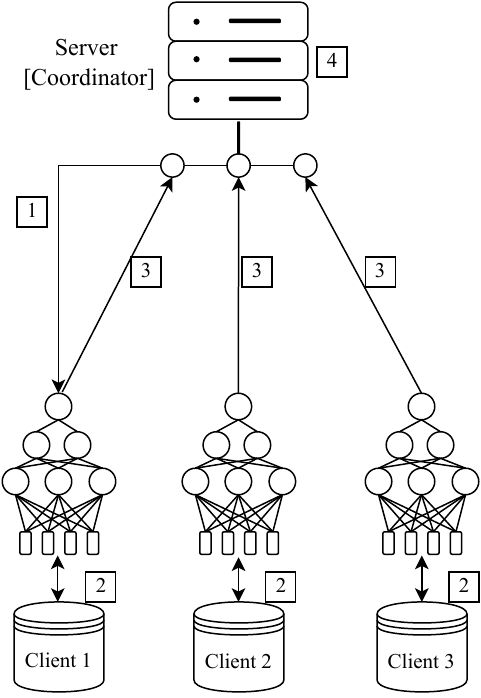}
         \caption{Decentralized/Federated.}
         \label{fig:decentralized}
     \end{subfigure}
     \caption{Architectures for machine learning.}
        \label{fig:comparisonCentralizedDeccentralized}
\end{figure}

The dynamic nature of \ac{FL} makes it challenging to present a complete overview of all its variants. In the following, we consequently focus on the conceptual differences between \ac{FL} and more conventional or centralized approaches to the training of \ac{ML} models on cross-organizational data (Figure~\ref{fig:comparisonCentralizedDeccentralized}).

In conventional or centralized approaches, \ac{ML} models are trained on (anonymized) data stored in a central repository. In more technical terms, the "learning" process proceeds as follows:

\begin{enumerate}
\item Clients send their data (usually anonymized) to a central repository.
\item The central repository translates this data into a pre-defined format.
\item A pre-defined model is trained on the pre-processed data.
\item The trained models are sent back to the clients that submitted their data.
\end{enumerate}

In the case of \ac{FL}, each client instead maintains control over its data set. It is thus also responsible for the preparation of this data for processing. As soon as the data is prepared, the clients must establish a process to coordinate and streamline the training of the \ac{FL} model. Once all clients have agreed to this process, they can start locally training a (partial) \ac{ML} model while maintaining full control of their data. When local training is completed, the locally trained models are aggregated and shared with the other clients. Usually, the aggregation of (partial) \ac{ML} models occurs on a central server. In more technical terms, the steps of this alternative "learning" process are as follows:

\begin{enumerate}
    \item A subset of clients is selected and downloads the most recent version of the "global" \ac{ML} model (typically from a central server).
    \item The clients train an updated, "local" model using their local data.
    \item The clients upload their updated local models (typically to the central server).
    \item The local models are aggregated to build a better "global" model, and the previous four steps are repeated.
\end{enumerate}

Multiple algorithms have been developed to select clients in each training round and aggregate their updated models to overcome problems with model heterogeneity, stability, and synchronization \citep[see][]{tlr_zhang2021survey}. The fundamental process, however, is always the same: \ac{ML} models are trained locally, shared with the other clients or a central server, aggregated, and then re-routed to clients for another training round. These rounds are repeated until a certain performance benchmark or a set number of training rounds has been reached.

\section{Organizational Opportunities}
\label{sec:org}

\begin{figure}[h!]
    \centering
    \includegraphics[scale=0.5]{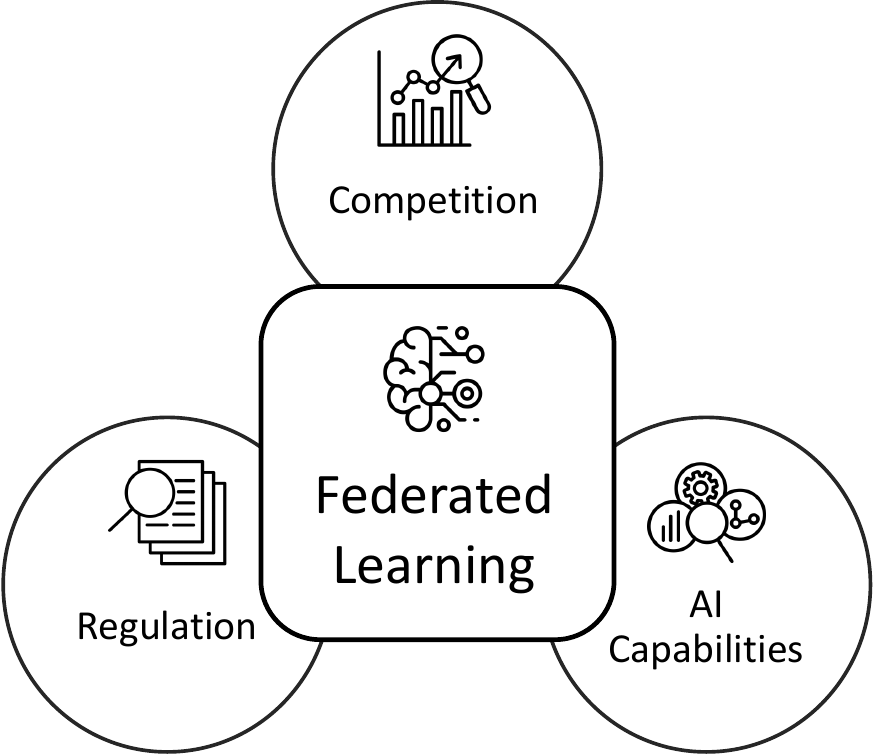}
    \caption{Federated learning at the intersection of regulation, competition, and AI capabilities.}
    \label{fig:triangle}
\end{figure}

Attractive \ac{FL} applications will typically be situated in a triangle between regulation, competition, and AI capabilities (see Figure\ref{fig:triangle}). While data regulation and competition limit the degree of (desirable) data sharing, \ac{AI} capabilities define the degree to which organizations can train, deploy, and use \ac{ML} models \citep{mikalef2021artificial}. 

\subsection{Regulation}
\label{subsec:regulation}

A first organizational dimension for the attractiveness of \ac{FL} relates to the prevalence of data-sharing regulations. In Europe, for example, the \ac{GDPR}, the Data Act, the Data Governance Act, and the proposed \ac{AI} Act strictly regulate how organizations can share data across organizational boundaries. For organizations required to comply with these regulations, \ac{FL} can be valuable as it avoids data sharing. The resulting "increase" in training data may also reduce the bias of the final \ac{FL} model \citep{debrusk2018bias, kostick2022mitigatingbias}. Multiple supervisory authorities, including sections of the European Data Protection Supervisor, thus voice support for \ac{FL} as a means to advance the causes of decentralization, data minimization, and (international) cooperation \citep{edps2023federatedlearning}.

\subsection{Competition}
\label{subsec:competition}

A second organizational dimension for the attractiveness of \ac{FL} is the prevalence of data-based competition. Organizations in such environments are often reluctant to share data, especially when data control goes along with control over data network effects \citep{Gregory2021,abbas2021business,levine2023networkeffects}. As \ac{FL} enables individual organizations to maintain control over their data, it can facilitate cooperation between otherwise competing organizations. Assuming well-defined governance and a fair allocation of surpluses, \ac{FL} may be especially attractive for smaller organizations, as they can use collaboratively trained \ac{ML} models to compete with other, larger organizations \citep{bammens2023scale, Zhang2023smemanufacturer}. However, \ac{FL} can also be useful for organizations that aim to promote de-facto standards for the use of \ac{ML} models in their industry~\citep{blasch2021standards}.

\subsection{AI capabilities}
\label{subsec:capabilities}

A third dimension that organizations interested in \ac{FL} should consider relates to their abilities to design, train, and run \ac{ML} models. Despite the increasing prevalence of open-source \ac{ML} models and AI-as-a-service offerings \citep{lins2021artificial}, many organizations struggle with the capabilities required to make use of these models and offerings \citep{lins2021artificial, Enholm2021}. \ac{FL} can alleviate some of these challenges by not only pooling data but also the capabilities that organizations must have in order to design, train, and run advanced \ac{ML} models. Especially for organizations with weaker technological capabilities, it can make sense to partner with and learn from stronger organizations \citep{zahoor2022alliance, easterby2008knowledge}. Yet also those with stronger technological capabilities can benefit when partner organizations can add superior business capabilities \citep{akter2016dc, henderson1990plugging}.

\section{Adoption Challenges}
\label{sec:adoption}

\begin{figure}[h!]
    \centering
    \includegraphics[scale=0.8]{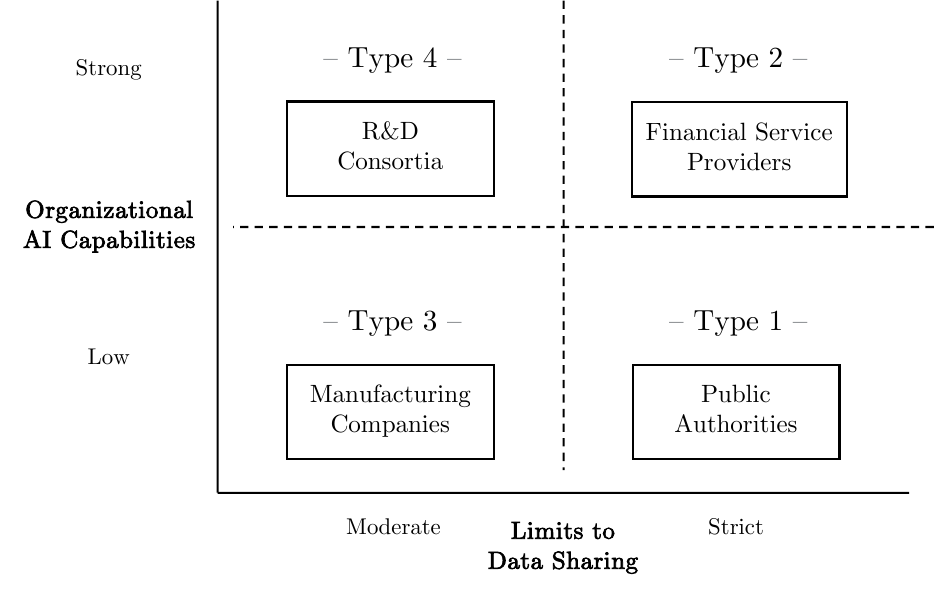}
    \caption{A conceptual framework for the adoption of FL}
    \label{fig:quadrant_empty}
\end{figure}

Building on the organizational opportunities in Section \ref{sec:org}, we now turn to the adoption challenges associated with \ac{FL}. To structure our discussion of these challenges, we employ a simple framework to distinguish organizations interested in deploying \ac{FL} according to their (regulatory and competitive) limits related to data sharing and the level of their \ac{AI} capabilities (see Figure \ref{fig:quadrant_empty}). Most of the presented challenges are present in all \ac{FL} projects, but some will be more pronounced for each "type" in our framework.

\subsection{Type 1 - Low AI capabilities \& Strict limits to data sharing}

Type 1 organizations are characterized by comparatively weak \ac{AI} capabilities and operate in environments with strict limits on data sharing. Examples of Type 1 organizations include healthcare providers and public authorities. These organizations may benefit the most from \ac{FL} as it enables them to "pool" previously isolated data silos \citep{balta2021flgov} and build better AI capabilities together \citep{sprenkampfederated, rieke2020future}.
 
However, they may also have the highest hurdles. Their legacy IT systems will sometimes not be able to easily provide the required training data ~\citep{bisbal1999legacy}. The integration of disparate data sources and formats from different organizations may pose significant data preprocessing and harmonization difficulties~\citep{bammens2023scale}. Many will also face challenges in ensuring data privacy and security during the FL process \citep{rieke2020future}.  These challenges can lead to suboptimal models and complicate the joint building of new AI capabilities\citep{kang2018organizational}.

\subsection{Type 2 - Strong AI capabilities \& Strict limits to data sharing}

Type 2 organizations are characterized by comparatively strong \ac{AI} capabilities and relatively strict limits to data sharing. Examples of these organizations include financial service providers that already use \ac{ML} to support various business functions and services \citep{desai2023financeai}. Using \ac{FL}, they can collectively train models both between organizational units that are not permitted to share data and enter strategic alliances with other organizations. 

Entering these alliances, however, may raise antitrust questions, especially when large organizations collaborate, and the overall alliance has a dominant market position. One way of addressing these questions could be to open-source the final model~\citep{eu2021commission, veale2021aiact, rab2019artificial}. However, it is unclear how antitrust authorities will respond to \ac{FL} alliances \citep{mahari2021antitrust}.

\subsection{Type 3 - Low AI capabilities \& Moderate limits to data sharing}

Type 3 organizations are characterized by comparatively low \ac{AI} capabilities and moderate limits on data sharing. They stand to benefit from \ac{FL} as it allows them to pool their limited \ac{AI} capabilities with other organizations facing similar challenges \citep{uberlee_2023}. Examples of Type 3 organizations include traditional manufacturing companies.

For these manufacturing companies, the long-term success of \ac{FL}-driven collaborations hinges on appropriate governance structures \citep{mit2023governance}. Setting up these structures will commonly involve organizational, economic, and legal considerations. As the AI capabilities among players can vary, preventative measures should be taken to avoid the exploitation of weaker participants~\citep{rodrigues2020legal}. With regards to the legal dimensions, clear agreements must be made that adequately address concerns about intellectual property and shared accountability for the final \ac{ML} model ~\citep{Bhaduri2003IntellectualPM}.

\subsection{Type 4 - Strong AI capabilities \& Moderate limits to data sharing}

Type 4 organizations have notable \ac{AI} capabilities and operate in environments with moderate limits to data sharing. They may profit from \ac{FL} when there are substantial benefits from highly accurate models but regulation complicates data sharing. Examples of Type 4 organizations include R\&D consortia, provided they do not have industry-specific restrictions on data- or model-sharing.

Despite moderate restrictions on data sharing, a central concern for Type 4 organizations is typically related to the security and protection of proprietary data. Although \ac{FL} promises a high degree of security and data protection, it cannot guarantee it ~\citep{benmalek2022security}. More specifically, the security of \ac{FL} implementations can be compromised by attacks targeting the communication between clients \citep{chatterjee2020applied,wang2020man} or by reverse-engineering training data from model updates \citep{bagdasaryan2020backdoor,shejwalkar2021manipulating}.

To mitigate these security risks, client communication can be safeguarded using cryptographic techniques such as\ac{SMPC} \citep{ben-or1988smpc, cramer2001mpc} or zero-knowledge proofs \citep{ruckel2022privacyincentives}. However, these methods are often computationally demanding and require additional capabilities \citep{zhao2019smpc}. The reverse-engineering of training data, which remains an open issue for \ac{SMPC}, can be avoided by adding noise using differential privacy. The implementation of differential privacy helps mitigate the potential for data leakage \citep{mugunthan2019smpai, xing2023zeroknowledge}. However, even with the recent advances, these techniques can sometimes lead to a decrease in model performance \citep{FERNANDEZ2022119915, kaissis2021end, mcmahan2018}.

\section{Research Opportunities}
\label{sec:discussion}

Machine learning projects often face significant challenges due to regulatory or competitive limits to data sharing. \ac{FL} provides a promising solution for these projects because it enables organizations to collaboratively train \ac{ML} models without sharing training data directly, fostering a more privacy-preserving environment and unlocking new opportunities for innovation and problem-solving in cross-organizational \ac{ML} contexts. In what follows, we discuss opportunities for Information Systems research to support the realization of these \ac{FL} opportunities and the resolution of the challenges we identified in Section \ref{sec:adoption}.

The first set of research opportunities pertains to the \textit{technical aspects} and applications of \ac{FL} \citep[see][]{lins2021artificial}. Despite continuous improvements, there are still limitations relating to security, privacy, and performance. More technical IS research can help identify these limitations, evaluate potential solutions, and develop guidance for aligning the technical aspects of \ac{FL} with the requirements of cross-organizational collaboration. Exemplary research questions could include:

\begin{itemize}
    \item What are best practices to ensure security and privacy in a \ac{FL} setting?
    \item Are there certain model architectures that better lend themselves to the use of \ac{FL}?
    \item How can \ac{FL} algorithms be optimized to handle heterogeneous and distributed data sources commonly found in organizational environments?
\end{itemize}

The second set of research opportunities pertains to the \textit{organizational aspects} in which \ac{FL} is used and how \ac{AI} is managed \citep[see][]{berente2021managing}. As \ac{FL} projects can bring together competing organizations with different AI capabilities, effective governance is highly important. While we expect that many insights from research on the outsourcing of \ac{IT} capabilities and consortia/partnership governance will also hold for \ac{FL}, existing governance frameworks might have to be expanded to address the technical details of \ac{FL} operations ~\citep{tlr_zhang2021survey, bammens2023scale}. Along these lines, possible research questions are:

\begin{itemize}
    \item What is the minimum level of data needed to create meaningful value through \ac{FL}?
    \item How can the value co-created through \ac{FL} be distributed fairly?
    \item What governance structures are needed for (decentralized) \ac{FL}?
    \item How can managers prevent conflicts of interest and promote the equitable participation of all stakeholders?
\end{itemize}

The third set of research opportunities pertains to the \textit{regulatory aspects} of specific \ac{FL} implementations\citep[see][]{TRUONG2021102402}. Although \ac{FL} enables compliance with data sharing regulations, its use creates new regulatory uncertainties such as in the area of antitrust. While many regulators and supervisory authorities, including the European Data Protection Supervisor, appear to be positive about \ac{FL}, it remains to be seen if they will further regulate \ac{FL}. Along these lines, possible research questions are:

 \begin{itemize}
     \item How can legal frameworks effectively balance data privacy regulations and innovation in cross-jurisdictional \ac{FL} deployments?
     \item How can regulators create a solid legal groundwork for public authorities to engage in \ac{FL} projects?
     \item How can legal safeguards and transparent data usage mechanisms be established to maintain data integrity, while addressing antitrust concerns and ensuring accountability within \ac{FL} collaborations?
\end{itemize}

\section{Conclusion}
\label{sec:outlook}

Unpacking the catchword "federated learning", we describe how strict limits on data sharing can be resolved through the collaborative training of \ac{ML} models without sharing their underlying data sets. We explain the technical foundations of \ac{FL} and how they differ from conventional or centralized \ac{ML} approaches. Building on these differences, we describe three dimensions that influence the organizational attractiveness of \ac{FL}. We then turn to adoption challenges and the manifold opportunities for IS research to support with resolving these challenges. 

Ultimately, \ac{FL} might have the potential to make \ac{AI} capabilities available to a broader group of organizations. It may also present interesting opportunities for more sustainable training of \ac{ML} models by shifting training rounds among the involved organizations in line with the availability of cheap and renewable power~\citep{anthony2020carbontracker,grange2018green}.

\section*{Credit authorship contribution statement}
Conceptualization, J.D.F, T.B; Methodology, J.D.F, T.B, A.R; Writing - Original Draft, J.D.F, T.B, M.B.;  Supervision A.R, G.F.; Writing - Review \& Editing, A.R, G.F.; Visualization, J.D.F; Funding acquisition, G.F. All authors have read and agreed to the published version of the manuscript.

\section*{Declaration of Competing Interest}
The authors declare no conflict of interest.

\section*{Acknowledgements}

We would like to acknowledge Chul Min Lee for his valuable feedback on the first draft of this paper.

Moreover, we would like to acknowledge the support of the European Union (EU) within its Horizon 2020 programme, project MDOT (Medical Device Obligations Taskforce), Grant agreement 814654. Additionally, supported by PayPal together with the Luxembourg National Research Fund FNR (P17/IS/13342933/PayPal-FNR/Chair in DFS/Gilbert Fridgen).

\bibliographystyle{IEEEtranN}
\bibliography{literature}  

\end{document}